\documentclass[aps,twocolumn,amsmath,amssymb,superscriptaddress,floatfix]{revtex4-1}

\usepackage{graphicx,xcolor}
\usepackage{soul}
\usepackage{ulem}
\usepackage{amsfonts,amssymb,amsmath}

\begin{document}

\title{Dimensional crossover of charge order in IrTe$_2$ with strong interlayer coupling}

\author{Hyoung Kug Kim}
\affiliation{Department of Physics, Pohang University of Science and Technology (POSTECH), Pohang 37673, Korea}

\author{So Young Kim}
\affiliation{Department of Physics, Pohang University of Science and Technology (POSTECH), Pohang 37673, Korea}
\affiliation{Department of Materials Science and Engineering, POSTECH, Pohang 37673, Korea}
\affiliation{Center for Artificial Low Dimensional Electronic Systems, Institute for Basic Science, Pohang 37673, Korea}

\author{C.~J.~Won}
\affiliation{Laboratory for Pohang Emergent Materials, POSTECH, Pohang 37673, Korea}
\affiliation{MPPC-CPM, Max Planck POSTECH/Korea Research Initiative, Pohang 37673, Korea}
\affiliation{Department of Physics, Pohang University of Science and Technology (POSTECH), Pohang 37673, Korea}

\author{Sang-Wook Cheong}
\affiliation{Laboratory for Pohang Emergent Materials, POSTECH, Pohang 37673, Korea}
\affiliation{MPPC-CPM, Max Planck POSTECH/Korea Research Initiative, Pohang 37673, Korea}
\affiliation{Rutgers Center for Emergent Materials and Department of Physics and Astronomy, Rutgers University, Piscataway, NJ, USA}

\author{Jonghwan Kim}
\affiliation{Department of Materials Science and Engineering, POSTECH, Pohang 37673, Korea}
\affiliation{Department of Physics, Pohang University of Science and Technology (POSTECH), Pohang 37673, Korea}
\affiliation{Center for vdW quantum solids, Institute for Basic Science, Pohang 37673, Korea}

\author{Jun Sung Kim}
\affiliation{Department of Physics, Pohang University of Science and Technology (POSTECH), Pohang 37673, Korea}
\affiliation{Center for Artificial Low Dimensional Electronic Systems, Institute for Basic Science, Pohang 37673, Korea}

\author{Tae-Hwan Kim}
\email{taehwan@postech.ac.kr}
\affiliation{Department of Physics, Pohang University of Science and Technology (POSTECH), Pohang 37673, Korea}
\affiliation{MPPC-CPM, Max Planck POSTECH/Korea Research Initiative, Pohang 37673, Korea}

\begin{abstract}
Tuning dimensionality in van der Waals materials with finite interlayer coupling has introduced various electronic phase transitions by conventional mechanical exfoliation.
Particularly when the electronic order is tied to the modulation of the interlayer coupling, such dimensional tunability has a strong impact on its stability and properties, which has rarely been investigated experimentally.
Here, we demonstrate a dimensional crossover of charge order in IrTe$_2$ from genuine two- to quasi-three-dimension using low-temperature scanning tunneling microscopy and spectroscopy.
Employing atomically thin IrTe$_2$ flakes ranging from monolayer to multilayer, we observe a gradual phase transition of charge order and exponential decay of Coulomb gap with increasing thickness. 
Moreover, we find a
suppression of the density of states emerging at an abrupt lateral interface between two- and three-dimension.
These findings are attributed to the interplay between the strongly coupled layers and substrate-driven perturbation, which can provide a new insight into the dimensional crossover of strongly coupled layered materials with hidden electronic phases.
\end{abstract}

\maketitle

\section{Introduction}
Dimensionality in a quasi-two-dimensional (quasi-2D) system works as one of critical physical parameters in determining structural and electronic characteristics,
which often induces various phase transitions including superconductivity~\cite{Xi2016, Song2021b}, charge density waves~\cite{Xi2015, Yu2015, Chen2022}, or symmetry breaking~\cite{Lin2017, Ni2021}.
Quasi-2D van der Waals (vdW) materials, such as transition metal dichalcogenides (TMDs), feature strong in-plane covalent bonding and weak out-of-plane vdW bonding, which enable tuning their dimensions by adjusting thickness without introducing disorder or defects~\cite{Ghosh2010, Jin2013}.
Because a typical electronic order in vdW materials is three-dimensional (3D) due to finite interlayer coupling~\cite{Yang2012, Jin2013, Yang2014}, modulating out-of-plane interaction can generate fascinating layer-dependent physical properties particularly when the electronic order is strongly linked to the modulation of interlayer bonding or stacking configuration~\cite{Lee2019, Lee2021}.

IrTe$_2$, one of exotic TMDs showing intertwined charge ordering phases, is associated by the structural phase transition from high temperature trigonal to low temperature monoclinic lattice ($T_{c} \sim 275$~K) with the wave vector of $q_{0}=(1/5,0,1/5)$~\cite{Matsumoto1999, Yang2012}.
Due to the intralayer Ir-Ir dimerization competing against the interlayer Te-Te bonding~\cite{Machida2013, Joseph2013, Dai2014}, the so-called cross-layer charge ordering with stripe charge modulations [$q_n = (3n+2)^{-1}$] have been observed in the low temperature phase~\cite{Hsu2013, Pascut2014}.
Furthermore, bulk intrinsic superconductivity emerges when the charge ordering is fully suppressed by chemical doping~\cite{Yang2012,Pyon2012,Ootsuki2012, Kamitani2013} or rapid cooling~\cite{Kim2016, Oike2018a}.
In contrast to bulk IrTe$_2$, the coexistence of superconductivity and stripe charge modulation 
has recently been observed in mechanically exfoliated flakes with the thickness range of 20--200 nm~\cite{Park2021}.
Such a contrasting observation suggests that the coexisting stripe charge order significantly enhances the out-of-plane coherence length and the coupling strength of superconductivity in thin IrTe$_2$ flakes. 
In this respect, IrTe$_2$ can be a proper model system for studying the role of interlayer coupling in the interplay between charge ordering and superconductivity at different dimensions, which has not been investigated experimentally.

In this article,
we have investigated atomically thin IrTe$_2$ flakes ranging from monolayer to multilayer (up to 20~L) in order to correlate a dimensional crossover of charge order with thickness.
Using the Al$_2$O$_3$-assisted mechanical exfoliation~\cite{Deng2018}, we systematically reduce the thickness of IrTe$_2$ down to monolayer and directly probe structural and electronic properties with low temperature scanning tunneling microscopy and spectroscopy (STM/STS) measurements.
We observe a gradual evolution of charge order and an exponentially decay of Coulomb gap, which have been attributed to the subtle interplay between the strong interlayer coupling and substrate-induced disorders.
Furthermore, we discover an unusual
depression of density of states (DOS) in empty states,
which appears at an abrupt lateral interface from 2D to 3D.
Our findings demonstrate that the stability of the cross-layer charge ordering and the phase competition with different ordering periods are highly sensitive to the dimensional crossover, highlighting the important role of the interlayer coupling in IrTe$_2$.

\begin{figure}[tbp]
\centering \includegraphics[width=0.98\linewidth]{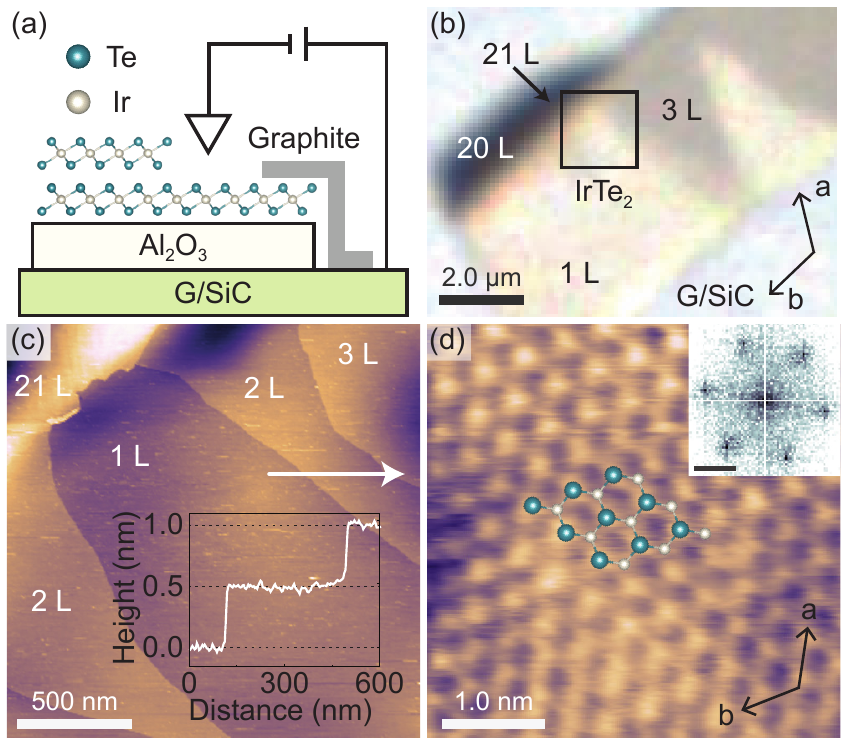}
\caption{\label{fig:1}
(a) Cross-sectional schematic of Al$_2$O$_3$-assisted exfoliated IrTe$_2$ onto epitaxially grown graphene on a single-crystalline silicon carbide substrate (G/SiC).
A thin exfoliated graphite flake is used for an electrical contact for STM/STS measurements. 
(b) Representative optical microscopy (OM) image of an atomically thin IrTe$_2$ flake with thickness ranging from 1~L up to 21~L.
(c) STM topographic image of the same flake obtained at a location which is indicated by a black box in (b).
Inset shows a STM height profile taken along a white arrow. 
Imaging conditions: $V_{\rm b}$ = 0.5~V; $I_{\rm t}$ = 30~pA; $T$ = 300~K. 
(d) Atom-resolved STM image of monolayer IrTe$_2$ showing a triangular lattice of topmost Te atoms.
Inset shows the fast Fourier transform (FFT) of the image, exhibiting only six atomic Bragg peaks without any charge order (scale bar, 2~nm$^{-1}$).
Imaging conditions: $V_{\rm b}$ = 3~mV; $I_{\rm t}$ = 5~nA; $T$ = 300~K.
}
\end{figure}

\begin{figure*}[tbp]
\centering \includegraphics[width = 0.98\linewidth]{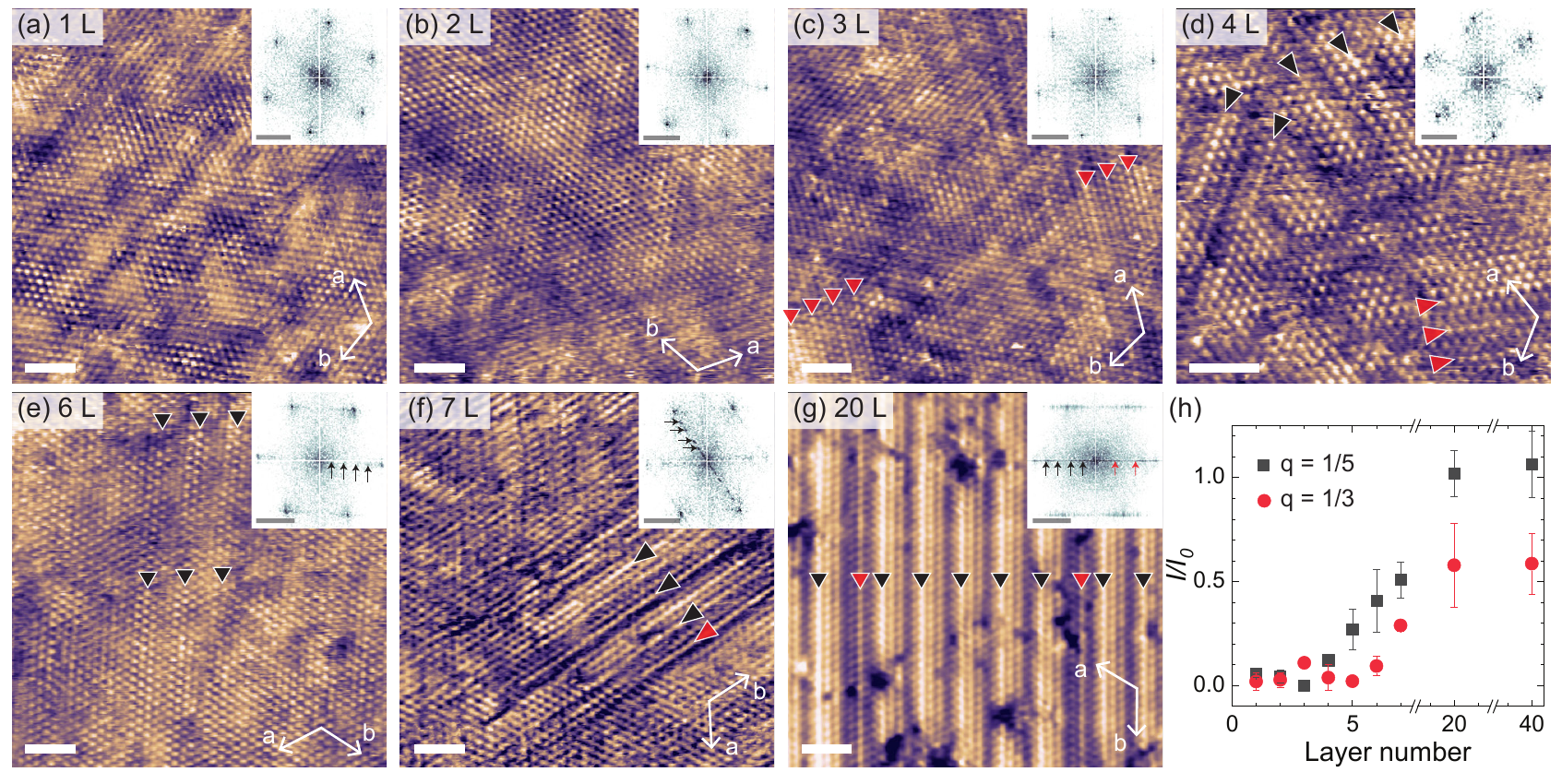}
\caption{\label{fig:2}
Atom-resolved STM images obtained on surfaces of IrTe$_2$ flakes with different layer numbers: (a) monolayer (1~L), (b) bilayer (2~L), (c) trilayer (3~L), (d) tetralayer (4~L), (e) hexalayer (6~L), (f) heptalayer (7~L), and (g) twenty-number layer ($V_{\rm b}$ = 4--200~mV; $I_{\rm t}$ = 1.0--6.0~nA; $T$ = 88~K; scale bar, 2~nm).
Red and black triangles indicate charge orders with the periodicity of $3a_0$ and $5a_0$, respectively.
Insets show the corresponding FFT images with black (red) arrows indicating $q = 1/5$ ($q = 1/3$) charge order (scale bar, 2~nm$^{-1}$).
(h) Relative ratio ($I/I_0$) of averaged intensity ($I$) of the charge order with either $q = 1/5$ (black squares) or 1/3 (red circles) with respect to Bragg peak intensity ($I_0$) in FFT images as a function of layer number.
}
\end{figure*}

\begin{figure*}[tbp]
\includegraphics[width = 0.98\linewidth]{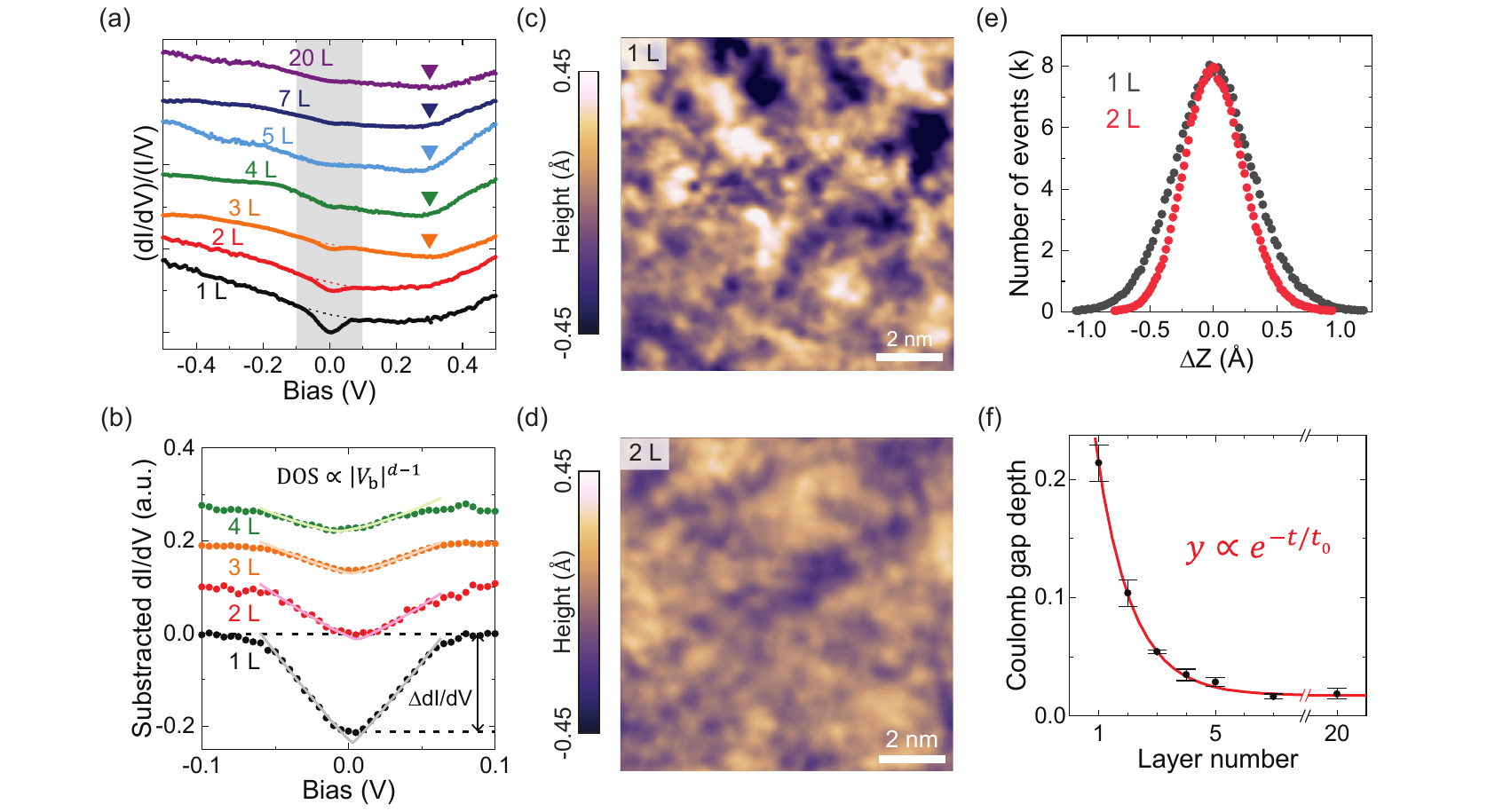}
\caption{\label{fig:3}
Thickness-dependent electronic properties of IrTe$_2$ flakes.
(a) Differential conductance spectra obtained on IrTe$_2$ flakes as a function of thickness.
All spectra are normalized and offset for clarity.
(b) Zoom-in spectra of (a) after the slow-varying background subtraction.
The light-colored Coulomb gap equation fits to the subtracted $dI/dV$ curves.
(c,d) STM topographic images of (c) 1~L and (d) 2~L IrTe$_2$.
The images are low-pass filtered to enhance surface roughness.
(e) Height histograms of (c) and (d). 
The standard deviation of 1~L is 33\% larger than that of 2~L.
(f) Coulomb gap depth as a function of layer number.
The depth is fitted by an exponential decay function with an effective screening length of $t_0 = 1.2$.
}
\end{figure*}

\begin{figure*}[tbp]
\centering \includegraphics[width = 0.98\linewidth]{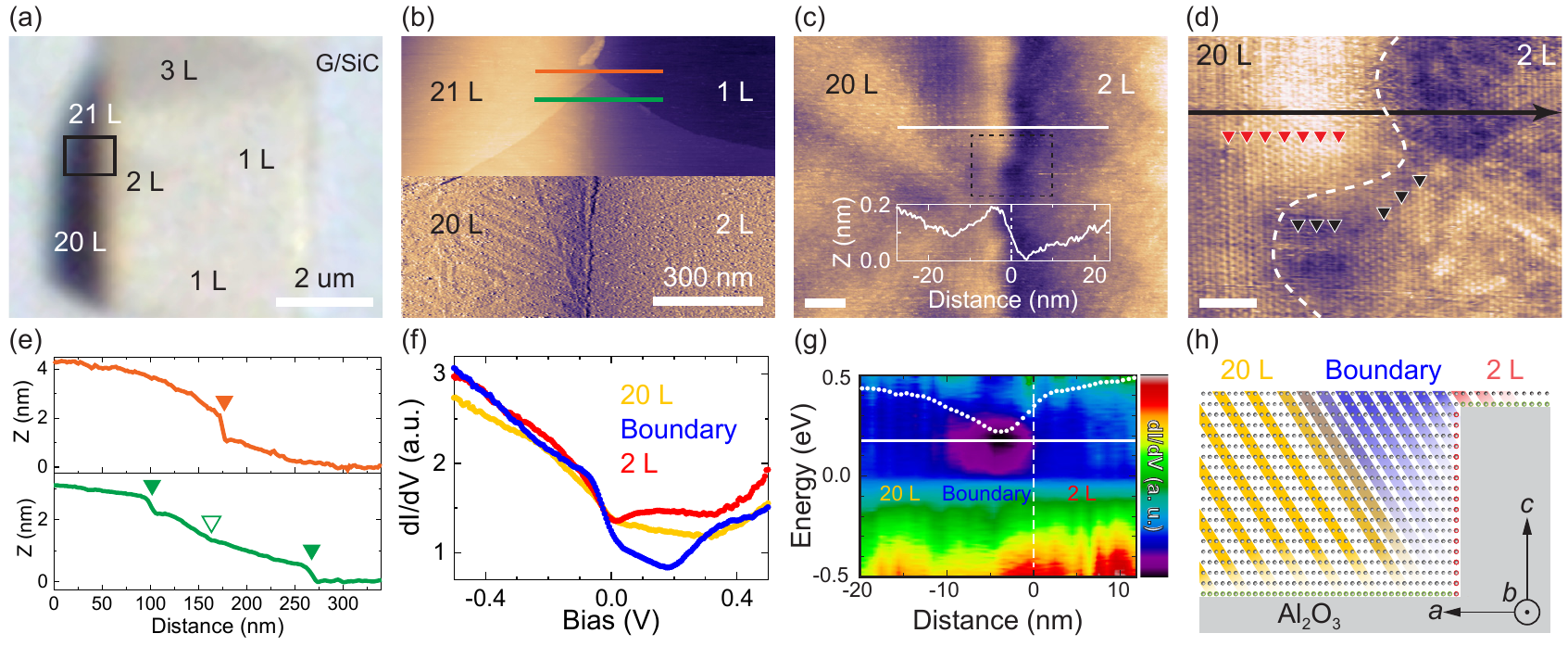}
\caption{\label{fig:4}
Atomically seamless 3D-2D interface. 
(a) OM image of an IrTe$_2$ flake including an atomically seamless 20/2~L interface.
(b) Large-scale STM topographic image of the 20/2~L interface indicated by a black rectangle in (a).
For clarity, the lower half shows differentiated STM topograph along the fast scan direction.
(c) Zoom-in STM topographic image of the 20/2~L interface ($V_{\rm b}$ = 100~mV; $I_{\rm t}$ = 2.5~nA; scale bar, 10~nm).
The inset shows a STM height profile along a white line.
(d) Atom-resolved STM image indicated by a black dashed rectangle in (c) ($V_{\rm b}$ = 10~mV; $I_{\rm t}$ = 5 nA; scale bar, 1~nm).
Red and black triangles denote the $q=1/3$ charge orders on 20~L and 2~L, respectively.
(e) STM height profiles along orange and green lines in (b).
The atomically seamless 20/2~L interface shows a gradual height change (open green triangle) while structral steps exhibit abrupt height changes (closed triangles).
(f) Averaged $dI/dV$ curves corresponding to three different regions (20~L, 2~L, and the boundary).
(g) Spatially resolved STS obtained along the black arrow in (d).
A white dotted curve shows an energy cut at 0.175~eV along a white solid line.
(h) Schematic cross-sectional view of the 20/2~L interface showing only Ir atoms for simplicity.
Gray circles represent intact Ir atoms while red and green circles denote edge-terminating atoms and those in the bottom layer on Al$_2$O$_3$, respectively.
Orange, blue and red shaded regions indicate the cross-layer charge ordering between interlayer Ir-Ir dimerizations with $q = 1/5$, $1/3$ on 20~L and $1/3$ on 2~L, respectively.
Note that the white dashed curve in (d) and the white dashed  line in (g) denote the interface between 20~L and 2~L.
}
\end{figure*}

\begin{figure*}[tb]
\includegraphics[width = 0.95\linewidth ]{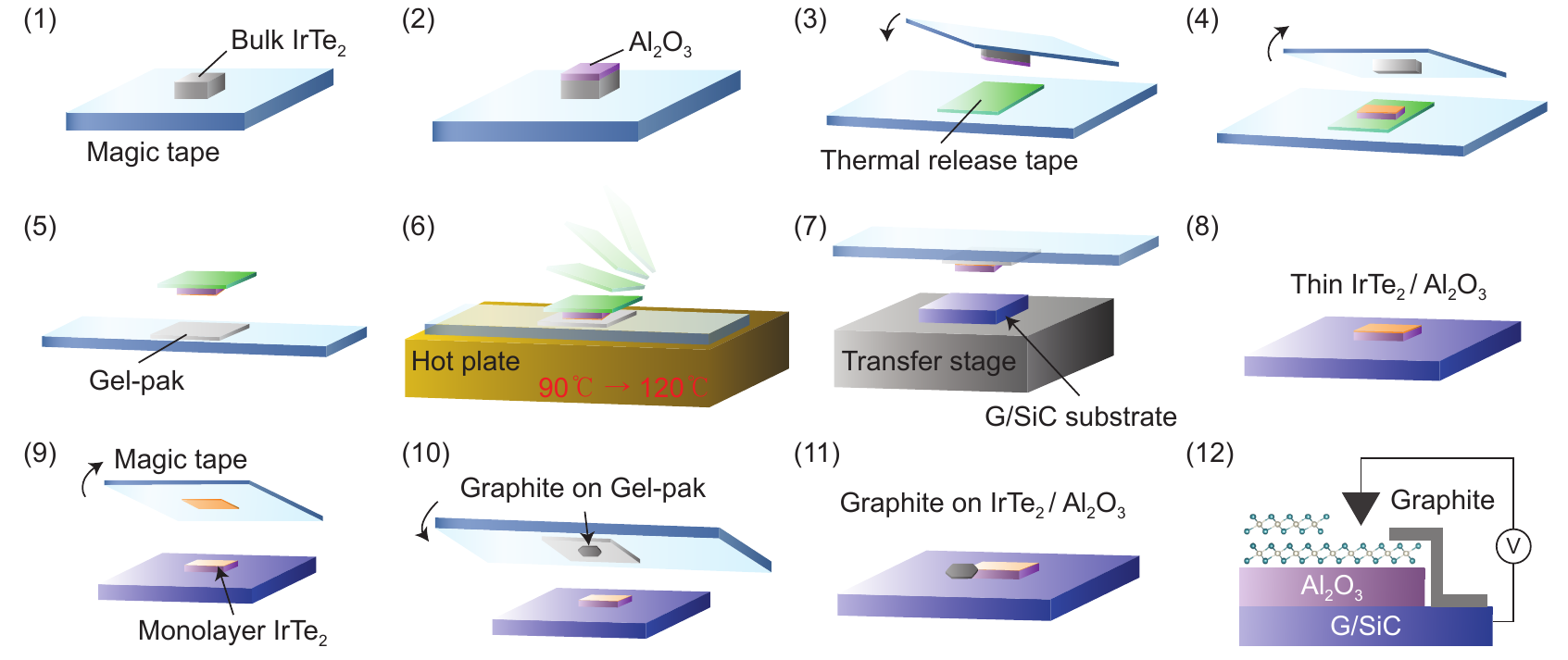}
\caption{Fabrication of atomically thin IrTe$_2$ flakes.
(1) Bulk IrTe$_2$ is freshly cleaved by the conventional Scotch tape exfoliation.
(2) Al$_2$O$_3$ is deposited on the freshly cleaved bulk IrTe$_2$ in vacuum.
(3,4) The flake is transferred onto a sheet of thermal release adhesive tape.
(5,6) The IrTe$_2$/Al$_2$O$_3$ flake is transferred onto a sheet of Gel-pak by heating the thermal release tape at 90--120\textcelsius.
(7,8) The flake is transferred onto graphene on a silicon carbide substrate (G/SiC).
(9) Atomically thin IrTe$_2$ flakes are produced by cleaving the flake with conventional Scotch tape.
(10) For an electrical contact, a piece of thin exfoliated graphite is transferred onto the target flakes.
(11) The sample is ready for STM measurements.
(12) Schematic drawing of our STM experimental setup.
}
\label{SM1}
\end{figure*}

\begin{figure*}[tb]
\centering \includegraphics[width = 1\linewidth]{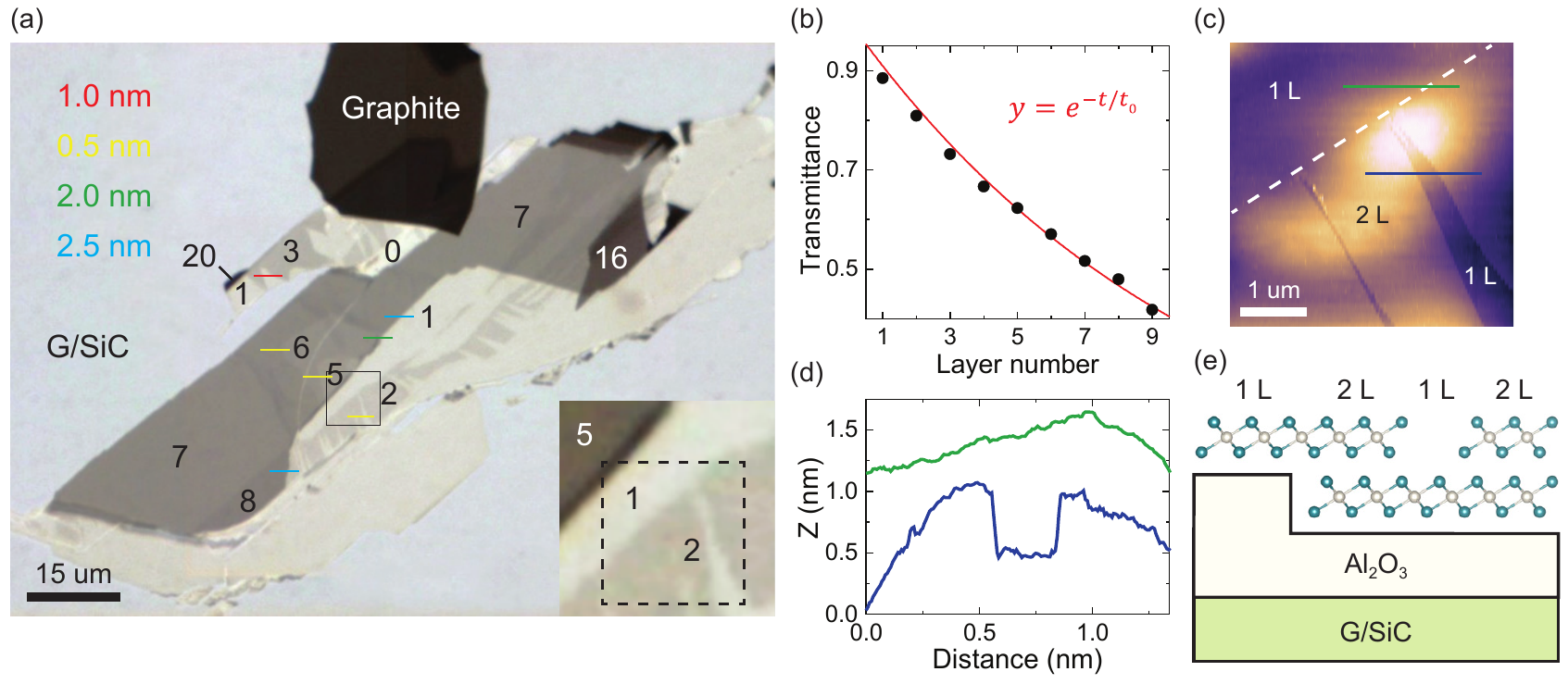}
\caption{\label{APP_fig_2}
Thickness determination and geometrical description of a IrTe$_2$/Al$_2$O$_3$ flake.
(a) OM image of a typical thin IrTe$_2$ flake with various thicknesses (1--20~L).
Numbers indicate the corresponding layer numbers of IrTe$_2$.
Colored horizontal lines denote height differences between neighboring regions from 0.5~nm to 2.5~nm.
The inset shows a zoom-in OM image indicated by a black square.
(b) Relative optical transmittance of IrTe$_2$ layers with respect to bare Al$_2$O$_3$ as a function of layer number
($t_0 = 10.5~\pm~0.1$).
(c) STM topographic image of a dashed square in the inset of (a).
The white dashed line indicates an atomic step underneath the surface.
(d) Height profiles across the 1/2~L interfaces indicated by a green and a blue line in (c).
The profiles are offset for clarity.
(e) Schematic cross-sectional view of monolayer and bilayer IrTe$_2$ in (c).
}
\end{figure*}

\section{Method}

Figures 1(a) and 1(b) show our experimental setup and a typical IrTe$_2$ flake fabricated by the Al$_2$O$_3$-assisted mechanical exfoliation in an inert atmosphere~\cite{Deng2018} (see more details in Appendix A).
We identify thicknesses of IrTe$_2$ using optical transmission contrasts, which are highly sensitive to layer numbers of IrTe$_2$ flakes (see Appendix B).
As shown in Figs.~1(b) and 1(c), we confirm the excellent correlation between the optical contrasts and STM height profiles.
Surface cleanliness, which is crucial for surface-sensitive STM,
is routinely checked by room-temperature STM measurements before cooling.
A typical atomic STM image of IrTe$_2$ monolayer clearly reveals the triangular Te atomic lattice without any significant surface contamination such as vacancies or adsorbates [Fig.~1(d)].
The sample was inserted to a cooled STM stage using a continuous flow cryostat.
STM/STS measurements were performed in an ultrahigh-vacuum ($8 \times 10^{-9}$~Pa) cryogenic STM ($T_{\rm STM}=88$~K).
All STM images were recorded in the constant-current mode with an electrochemically etched W tip.
STS measurements were performed by recording $dI/dV$ spectra via a lock-in technique with feedback open, using a modulation frequency of approximately 1~kHz with a root-mean-square modulation amplitude of $V_{\rm mod} = 7$~mV.


\section{Result and discussion}
To investigate thickness-dependent charge orders, we have performed STM measurements on atomically thin IrTe$_2$ flakes from monolayer (1~L) to multilayer up to 20~L well below the bulk transition temperature.
In sharp contrast to the dimerized monolayer and bilayer IrTe$_2$ grown by molecular beam epitaxy (MBE) on graphene~\cite{Hwang2022}, we found no signature of charge orders in our mechanically exfoliated monolayer and bilayer IrTe$_2$ [Figs.~2(a) and 2(b)].
Local stripe charge orders start to emerge from trilayer (3~L) and become dominant above
heptalayer (7~L) IrTe$_2$ [Figs.~2(c)--2(g)].
To quantify the thickness dependence of the stripe charge orders, we analyze the relative peak intensities of fast Fourier transform (FFT) of charge orders with respect to Bragg peaks from atom-resolved STM images [Fig.~2(h)]~\cite{Pasztor2017}.
To improve the reliability, we statistically averaged the relative peak intensities obtained from a sufficient number ($\sim$10) of topographic images on different locations for each thickness.
The charge order with the wave vector of $q_0 = (1/5,0,1/5)$ gradually becomes stronger with increasing thickness above 3~L while the $q = 1/3$ charge order exists locally on 3~L and becomes more visible above 5~L. 
Beyond 20~L, we found no further thickness dependence of the charge order as reported previously~\cite{Park2021}.
This observation strongly suggests that the stripe charge order
of few layer IrTe$_2$ exhibits a gradual evolution with increasing thickness, in contrast to the abrupt metal-insulator transition between MBE-grown monolayer and bilayer IrTe$_2$~\cite{Hwang2022}.

The structural distortion and strain effect from the substrate may give rise to thickness-dependent lattice constants, which may lead to the observed thickness dependence of the charge order.
To check this possibility, we have performed the quantitative analysis of the lattice constants on various thicknesses (1--20~L).
Within our experimental measurement error ($\pm$0.15~\AA), few-layer IrTe$_2$ on Al$_2$O$_3$ does not show significant thickness dependence of the lattice constants (3.65--3.82~\AA).
Thus, we can exclude that atomic distortion and strain effect directly from the Al$_2$O$_3$ substrate on the thickness dependence of the charge order.

To further understand the dimensionality-induced evolution of the charge orders, we systematically probe local DOS of IrTe$_2$ as a function of thickness by normalizing differential tunneling conductance [Fig.~3(a)].
All spatially averaged STS spectra on IrTe$_2$ flakes (1--20~L) show metallic behaviors. 
Such an observation is distinct from that in the MBE-grown monolayer IrTe$_2$ with a band gap larger than 1~eV~\cite{Hwang2022}.
While the normalized STS data exhibit a similar trend for all thicknesses, a close look reveals that an enhanced DOS appears above 0.3~eV (indicated by triangles) in IrTe$_2$ flakes thicker than bilayer.
This DOS enhancement is originated from the Ir-Ir dimerization in the charge-ordered phase, which is supported by density functional theory calculations~\cite{Pascut2014}.
Thus, this thickness-dependent spectroscopic feature is quite consistent with the observed gradual evolution of charge orders with increasing thickness.


In addition, we observe a significant DOS suppression [gray region in Fig.~3(a)] at the Fermi energy ($E_{\rm F}$).
To visualize more clearly the DOS suppression, we subtract quartic polynomial fits [dashed lines in Fig.~3(a)] from the normalized differential conductance spectra.
Each fit is obtained from the corresponding spectrum using data with 0.1~V$<|V_{\rm b}|<$0.2~V.
The subtraction results in the symmetric `V'-shaped DOS suppression that becomes more pronounced with decreasing thickness [Fig.~3(b)].
Such a series of suppressed DOS features are often observed in disordered metallic thin film due to disorder-enhanced Coulomb interaction, known as a Coulomb gap~\cite{Efros1975}.
This resemblance suggests that the underlying Al$_2$O$_3$ substrate causes electronic disorders in IrTe$_2$, which can lead to the thickness-dependent `V'-shaped DOS suppression.

As an atomically thin layer such as graphene monolayer simply follows the contours of the underlying substrates~\cite{Lui2009},
thin IrTe$_2$ flakes may also reflect surface corrugation of the underlying Al$_2$O$_3$ substrate.
To unveil such substrate induced effect, we carefully analyze the surface roughness of 1~L and 2~L IrTe$_2$ with low-pass filtered STM topographic images.
Figures~3(c) and 3(d) show the filtered STM topographic images of Figs.~2(a) and 2(b), respectively, by masking the outer six Bragg points from the atomic lattice.
The filtered images clearly show the thickness-dependent surface corrugation  [Fig.~3(e)] due to the underlying Al$_2$O$_3$.
The decrease in roughness with increasing thickness strongly indicates that the structural inhomogeneity is originated from the underlying Al$_2$O$_3$,
which naturally leads to the thickness-dependent electronic disorder causing the enhanced Coulomb interaction in IrTe$_2$ flakes.

A quantitative analysis of the Coulomb gaps demonstrates that the gap depth, 
which is defined by $\Delta dI/dV$ at $E_{\rm F}$ [Fig.~3(b)], exponentially decreases with increasing layer number, 
where the effective screening layer number is $t_0 = 1.2$ corresponding to only 0.6~nm [Fig.~3(f)].
The similar interlayer screening with an exponentially decaying behavior is also observed in few-layer graphene~\cite{Lee2009} and MoS$_2$ nanoflakes~\cite{Li2013} with thicker effective screening layer numbers of 10 and 3.5, respectively.
The much smaller screening layer number indicates that IrTe$_2$ has much high DOS at $E_{\rm F}$ than other vdW materials.
Since STM exclusively probes topmost surface layers, the observed exponential decay strongly suggests that metallic IrTe$_2$ overlayers effectively screen the Coulomb gap feature of the bottom layer, which is directly affected by the underlying Al$_2$O$_3$ substrate.

Another intriguing analysis provides a dimensional crossover of the system with the disorder-driven Coulomb gap.
By fitting the Coulomb gap equation~\cite{Efros1976} to 
the subtracted $dI/dV$ curves [Fig.~3(b)], we can obtain  
thickness-dependent dimensionalities. 
Since the effective energy resolution ($\Delta E$) of STM is given by $\Delta E \approx \sqrt{(3.5k_{\rm B}T)^2 + (2.5eV_{\rm mod})^2}$~\cite{Lauhon2001}, we exclude data points of $|V_{\rm b}|$$<$15.6~mV near $E_{\rm F}$ in the fitting process due to the energy resolution obtained from both thermal broadening ($T$ = 88~K) and a modulation amplitude ($V_{\rm mod}=7$~mV) of a lock-in amplifier in our experiments.
For 1~L IrTe$_2$, the DOS exhibits a linear dependence on bias voltage ($V_{\rm b}$).
According to the Coulomb gap equation (DOS $\propto |V_{\rm b}|^{d-1}$), we obtain the dimensionality $d = 2.02~\pm~0.1$ for 1~L.
This strongly indicates that 1~L IrTe$_2$ on Al$_2$O$_3$ behaves as genuine 2D with the disorder-induced Coulomb gap, which has mainly been investigated in theoretical approach~\cite{Lee2016, Szabo2020}.
With increasing thickness, we find that the DOS near $E_{\rm F}$ gradually loses its linear feature due to the strong interlayer coupling.
With the strong interlayer interaction, the topmost layer in thicker IrTe$_2$ can experience the Coulomb gap of the electronically disordered bottom layer, which leads to the resulting intermediate dimensionality between 2D and 3D.

Occasionally, we can find an atomically seamless interface between thin and thick IrTe$_2$ as shown in Figs.~4(a)--(e).
Such an interface would provide an unprecedented platform to investigate an interface between 2D and 3D without abrupt steps on the surface.
The atom-resolved STM image [Fig.~4(d)] and the height profile [Fig.~4(e)] across the 20/2~L interface clearly demonstrate that the topmost IrTe$_2$ layer is seamlessly connected between 20~L and 2~L while the 21/1~L interface shows an 1-nm-high atomic step.
As discussed above, the thicker 21~L and 20~L show the charge orders while the thinner 1~L and 2~L do not [Fig.~4(b)].
In addition, the atom-resolved STM image shows an unambiguous charge order with period of $3a_0$ near the interface as indicated by triangles in Fig.~4(d), which is somewhat different from the ground states in bulk showing the $q=1/6$ charge order~\cite{Oh2013, Dai2014, Chen2017}.

To understand the electronic property of the intriguing interface,
the spatially resolved STS measurement is taken across the interface along a black arrow in Fig.~4(d), which exhibits an apparent DOS suppression near $V_{\rm b}$ = 0.175~V mainly on the 20~L area with width of 10~nm from the interface [Fig.~4(g)].
Such a state near the interface can be qualitatively explained by the strong interlayer coupling. 
As illustrated in Fig.~4(h), the interlayer coupling delivers electronic properties of edge-terminating atoms (indicated by red circles) to the topmost layer by inducing an exotic DOS suppression only on the 20~L side.
On the other hand, the 2~L side does not exhibit such a DOS suppression because its top layer is not connected to any edge-terminating atoms.
This finding suggests that the strong interlayer coupling in IrTe$_2$ also plays a vital role in an atomically seamless interface with abrupt thickness changes.

Lastly, the observation of the ($3\times1$) charge orders locally near the 2~L region [Fig.~4(d)] contrasts to the absence of the stripe charge order in the isolated bilayer IrTe$_2$ [Fig.~2(b)].
It can be considered as a proximity effect due to the neighboring 20~L.
Remarkably, the signatures of the $q = 1/3$ charge order are found nearly $\sim10$~nm apart from the interface.
The long-range proximity effect in the metallic system is quite unusual, which may indicate the inherent instabilities of IrTe$_2$ against various stripe charge orders.
Particularly, the type and stability of the stripe charge order is found to be extremely sensitive to the strain of just $\sim0.1$\%~\cite{Nicholson2021}.
A possible strain near the interface can stabilize the $q = 1/3$ stripe charge ordering phase in the otherwise disorder-induced pristine phase.
Therefore, this finding highlights the incipient instability of IrTe$_2$ to charge ordering, which would provide an unprecedented way to create exotic charge-ordered phases by introducing local strains or terminating atomic edges with functional molecules~\cite{Zhao2018}.


\section{Conclusion}
We have fabricated the atomically thin and clean IrTe$_2$ flakes varying from monolayer to multilayer by utilizing the Al$_2$O$_3$-assisted exfoliation method,
which can overcome the strong coupling between Te interlayers.
Our STM/STS measurements exhibit the gradual evolution of charge orders and the exponentially decaying Coulomb gaps with increasing thickness.
Our finding reveals that these phenomena are originated from the interplay between the stripe charge ordering with strong interlayer coupling and the substrate-driven disorder.
Furthermore, the induced DOS suppression from edge-terminating atoms is found at the atomically seamless interface via the strong interlayer coupling.
Our observation strongly suggests that thinner IrTe$_2$ layers would have relatively much lower transition temperature of the charge ordering than our measurement temperature.
Further experiments are needed to determine the thickness dependence of the transition temperature.

\appendix
\section{Sample Fabrication}

To obtain atomically thin IrTe$_2$ flakes, we employ the Al$_2$O$_3$-assisted fabrication method~\cite{Deng2018} as shown in Fig.~5.
Contrary to the conventional tape exfoliation method~\cite{Park2021},
this method can produce atomically thin few-layer IrTe$_2$ down to monolayer.
For STM measurements, we need to use a piece of exfoliated graphite to bypass insulating Al$_2$O$_3$ between thin IrTe$_2$ flakes and a conducting substrate [see (12) in Fig.~5].
All the fabrication processes should be conducted in an Ar-filled glove box,
which prevents any unwanted oxidation or contamination due to oxygen or water.
Finally, atomically thin IrTe$_2$ flakes are transferred to an ultra-high vacuum STM chamber without any exposure to air through a home-built suitcase~\cite{Park2021}.

\section{Thickness Determination}
To determine the layer number of IrTe$_2$, we analyze optical transmittance of IrTe$_2$ layers according to the Beer-Lambert law~\cite{Zhang2017, Niu2018}.
As shown in Fig.~6(b), the quantized transmittance monotonically decreases with increasing layer number.
To assign layer numbers unambiguously, STM height measurements are required as demonstrated in Figs.~6(c)--(e).
For example, we expect a single atomic step at the 1/2~L interface marked by a dashed line from the OM transmittance contrast [inset of Fig.~6(a)].
However, the corresponding STM image [Fig.~6(c)] does not show any atomic step on the top surface, which means that the expected atomic step exists on the bottom side.
In this way, we unambiguously assign thicknesses of IrTe$_2$ flakes as shown in Fig.~6(e).


\begin{acknowledgments} 
This work was supported by the National Research Foundation of Korea (NRF) funded by the Ministry of Science and ICT, South Korea (Grants No. NRF-2021R1F1A1063263, 2021R1A6A1A10042944, 2022R1C1C2006027, and 2022M3H4A1A04074153).
SYK and JSK was supported by the Institute for Basic Science (IBS) through the Center for Artificial Low Dimensional Electronic Systems (no. IBS-R014-D1).
J.K. acknowledges the support from the NRF of Korea grants (NRF-2020R1A2C2103166).
SWC is funded in part by the center for Quantum Materials Synthesis (cQMS), funded by the Gordon and Betty Moore Foundation’s EPiQS initiative through grant GBMF10104, and by Rutgers University.
\end{acknowledgments}

%

\end{document}